\pdfoutput=1
\documentclass[longauth]{aa}
\usepackage{txfonts}
\usepackage{graphicx}
\usepackage{amssymb}
\usepackage{pslatex}
\usepackage{natbib}
\bibpunct{(}{)}{;}{a}{}{,}
\usepackage{color}
\usepackage{xspace}
\def\AA {{$\mathring{\rm A}$}\xspace}
\def\FB {F_\mathrm{B}}
\def\FS {F_\mathrm{S}}
\def\DS {D_\mathrm{S}}
\def\DL {D_\mathrm{L}}

\def\tE {t_\mathrm{E}}
\def\ThE {\theta_\mathrm{E}}

\def\RS {R_\mathrm{*}}
\def\ThS {\theta_\mathrm{*}}

\def\rhoS {\rho_\mathrm{*}}
\def\msun {M_{\odot}}

\def\rsun {R_{\odot}}

\def\teff {T_\mathrm{eff}}

\def\uo {u_{0}}
\def\to {t_{0}}

\def\VIo {(V-I)_{0}}

\def\VKo {(V-K)_{0}}
\def\JKo {(J-Ks)_{0}}
\def\HKo {(H-Ks)_{0}}
\def\JHo {(J-H)_{0}}
\def\Vo {V_{\mathrm{0}}}
\def\Io {I_{0}} 
\def\Ko {K_{0}}

\def\Kso {K_{\mathrm{s}, 0}}
\def\Ai {A_{\rm I}}

\def\Aks{A_{K_s}}
\def\AJ{A_J}
\def\AH{A_H}
\def\Ri {R_{\rm I}}
\def\EVI{E(V-I)}
\def\logg {\log g}

\def\d{\mathrm{d}}

\def\MiRC {M_{I,{\rm RCG}}}

\def\muas {\mu \mathrm{as}}

\newcommand\Eq[1]{Eq.~(\ref{#1})}
\newcommand\Fig[1]{Fig.~\ref{#1}}
\newcommand\Tab[1]{Table~\ref{#1}}
\newcommand\Sec[1]{Sect.~\ref{#1}}

\newcommand \dix[1] {\times 10^{#1}}
\def\event{OGLE~2004-BLG-482\xspace}

\def\badfit{{\,[\star]}}

\def\Apspl{A_{\rm pspl}}
\def\Aext{A_{\rm lld}}
\usepackage{ulem}
\usepackage{color}

\begin{document}
  \title{Limb-darkening measurements for a cool red giant in
    microlensing event OGLE 2004-BLG-482\thanks{Partly based on observations made at ESO (073.D-0575A).}}
  \titlerunning{Limb-darkening measurements for a cool red giant}
  \authorrunning{M.~Zub et al.}
  \author{
    M.~Zub\inst{1,2,3,4} \and A.~Cassan\inst{5,1,4} \and
    D.~Heyrovsk\'y\inst{6} \and P.~Fouqu\'e\inst{7,4} \and 
    H.C.~Stempels\inst{8} \and 
    M.D.~Albrow\inst{9,4} \and 
    J.-P.~Beaulieu\inst{5,4} \and S.~Brillant\inst{10,4} \and
    G.W.~Christie\inst{11,12} \and N.~Kains\inst{13,4} \and 
    S.~Koz{\l}owski\inst{14,12} \and D.~Kubas\inst{5,10,4} \and 
    J.~Wambsganss\inst{1,4}\and \\ and \\
    V.~Batista\inst{5} \and 
    D.P.~Bennett\inst{15} \and
    K.~Cook\inst{16} \and C.~Coutures\inst{5} \and
    S.~Dieters\inst{5} \and 
    M.~Dominik\inst{13,} \thanks{Royal Society University Research
    Fellow} \and 
    D.~Dominis~Prester\inst{17} \and 
    J.~Donatowicz\inst{18} \and 
    J.~Greenhill\inst{19} \and K.~Horne\inst{13} \and 
    U.G.~J{\o}rgensen\inst{20} \and S.R.~Kane\inst{21} \and 
    J.-B.~Marquette\inst{5} \and 
    R.~Martin\inst{22} \and J.~Menzies\inst{23} \and
    K.R.~Pollard\inst{9} \and K.C.~Sahu\inst{24} \and
    C.~Vinter\inst{20} \and A.~Williams\inst{22} \\
    (The PLANET Collaboration) \\
    A.~Gould\inst{14} \and D.L.~DePoy\inst{14} \and A.~Gal-Yam\inst{25} \and 
    B.S.~Gaudi\inst{14} \and C.~Han\inst{26} \and 
    Y.~Lipkin\inst{27} \and D.~Maoz\inst{28} \and 
    E.O.~Ofek\inst{29} \and B.-G.~Park\inst{30} \and 
    R.W.~Pogge\inst{14} \and J.~McCormick\inst{31}\and \\ 
    (The $\mu$FUN Collaboration) \\
    A.~Udalski\inst{32} \and M.K.~Szyma{\'n}ski\inst{32} \and
    M.~Kubiak\inst{32} \and G.~Pietrzy{\'n}ski\inst{32,33} \and
    I.~Soszy{\'n}ski\inst{32} \and 
    O.~Szewczyk\inst{32} \and {\L}.~Wyrzykowski\inst{32,34} \\
    (The OGLE Collaboration)
  }
  \institute{
    Astronomisches Rechen-Institut (ARI), Zentrum f\"{u}r Astronomie
    der Universit\"{a}t Heidelberg (ZAH), M\"{o}nchhofstr. 12­-14,
    69120 Heidelberg, Germany \and 
    International Max-Planck-Research
    School for Astronomy \& Cosmic Physics at the University of
    Heidelberg, Germany \and
    Institute of Astronomy, University of Zielona G\'ora, Lubuska
    st. 2, 65-265 Zielona G\'ora, Poland \and 
    The PLANET Collaboration \and 
    Institut d'Astrophysique de Paris, UMR~7095 CNRS -- Universit\'{e}
    Pierre \& Marie Curie, 98bis Bd Arago, 75014 Paris, France \and
    Institute of Theoretical Physics, Charles University, V
    Hole\v{s}ovi\v{c}k\'{a}ch 2, 18000 Prague, Czech Republic \and 
    LATT, Universit\'{e} de Toulouse, CNRS, 31400 Toulouse, France \and 
    Department of Physics and Astronomy, Box 516,                                 
    SE-751 20, Uppsala, Sweden \and                     
    University of Canterbury, Department of Physics \& Astronomy, 
    Private Bag 4800, Christchurch, New Zealand \and
    European Southern Observatory (ESO), Casilla 19001, Vitacura 19,
    Santiago, Chile \and
    Auckland Observatory, P.O. Box 24-197, Auckland, New Zealand \and
    The $\mu$FUN Collaboration \and
    University of St Andrews, School of Physics \& Astronomy, North
    Haugh, St Andrews, KY16~9SS, United Kingdom \and 
    Department of Astronomy, Ohio State University, 140 W. 18th Ave., 
    Columbus, OH 43210, USA \and 
    University of Notre Dame, Physics Department, 225 Nieuwland Science Hall, 
    Notre Dame, IN 46530, USA \and
    Institute of Geophysics and Planetary Physics, L-413, Lawrence Livermore 
    National Laboratory, P.O. Box 808, Livermore, CA 94550, USA \and
    Department of Physics, University of Rijeka, Omladinska 14, 51000
    Rijeka, Croatia \and
    Technical University of Vienna, Dept. of Computing, Wiedner 
    Hauptstrasse 10, Vienna, Austria \and
    University of Tasmania, Physics Department, GPO 252C, Hobart,
    Tasmania 7001, Australia \and
    Niels Bohr Institute and Centre for Star and Planet Formation,
    Juliane Mariesvej 30, 2100 Copenhagen, Denmark \and
    NASA Exoplanet Science Institute, Caltech, MS 100-22, 770 South 
    Wilson Avenue Pasadena, CA 91125, USA \and
    Perth Observatory, Walnut Road, Bickley, Perth 6076, Australia \and
    South African Astronomical Observatory, P.O. Box 9 
    Observatory 7935, South Africa \and
    Space Telescope Science Institute, 3700 San Martin Drive, 
    Baltimore, MD 21218, USA \and
    Astrophysics Group, Faculty of Physics, Weizmann Institute of Science, 
    Rehovot 76100, Israel \and
    Department of Physics, Institute for Basic Science Research, Chungbuk National 
    University, Chongju 361-763, Korea \and
    School of Physics and Astronomy and the Wise Observatory, Raymond and Beverly 
    Sackler Faculty of Exact Sciences, Tel-Aviv University, Tel Aviv 69978, Israel \and
    School of Physics and Astronomy, Tel-Aviv University, Tel-Aviv 69978, Israel \and
    Division of Physics, Mathematics and Astronomy, California Institute of 
    Technology, Pasadena, CA \and
    Korea Astronomy and Space Science Institute, Daejon 305-348, Korea \and
    Farm Cove Observatory, Centre for Backyard Astrophysics, Pakuranga, 
    Auckland New Zealand \and
    Warsaw University Observatory. Al. Ujazdowskie 4, 00-478 Warszawa, Poland \and
    Universidad de Concepci\'{o}n, Departamento de F\'{\i}sica, Casilla 160-C, 
    Concepci\'{o}n, Chile \and
    Institute of Astronomy, University of Cambridge, Madingley Road, 
    CB3 0HA Cambridge, UK
  }

  \date{ Received / Accepted }
  \abstract
      {}
      { 
	We present a detailed analysis of \event, a relatively 
	high-magnification single-lens microlensing event that exhibits 
	clear extended-source effects. These events are relatively rare,
	but they potentially contain unique information on the stellar atmosphere 
	properties of their source star, as shown in this study.
      }
      { 
	Our dense photometric coverage of the overall light curve and a proper microlensing
	modelling  allow us to derive measurements of the \event source star's 
	linear limb-darkening coefficients in three bands, including
	standard Johnson-Cousins $I$ and $R$, as well as in a broad clear filter. In particular, we
	discuss in detail the problems of multi-band and multi-site modelling on the expected
	precision of our results. We also obtained high-resolution UVES spectra as part of a ToO
	programme at ESO VLT, from which we derive the source star's precise fundamental parameters. 
      }
      { 
	From the high-resolution UVES spectra, we find that \event's source star is a red giant
	of MK type a bit later than M3, with $\teff = 3667 \pm 150$~K, $\logg = 2.1 \pm 1.0$ and
	an assumed solar metallicity. This is confirmed by an OGLE calibrated colour-magnitude diagram.
	We then obtain from a detailed microlensing modelling of the light curve linear
	limb-darkening coefficients that we compare to model-atmosphere predictions available in the
	literature, and find a very good agreement for the $I$ and $R$ bands. 
	In addition, we perform a similar analysis using
	an alternative description of limb darkening based on a principal component analysis of ATLAS
	limb-darkening profiles, and also find a very good agreement between measurements and model
	predictions. 
      }
      {}
  \maketitle

  \section{Introduction} \label{sec:intro}

  Photometric and spectroscopic observations of stars yield their spectral 
  types and other information useful for
  studying their atmospheres, but much of the information on the structure of the atmosphere
  and related physical processes is lost in the disc-integrated flux. 
  Advanced models calculated for a broad range of stellar types 
  (e.g. MARCS, \citet{Gustafsson2008}; ATLAS, \citet{Kurucz1992}; \citet{Plez1992})
  describe the corresponding physics at different optical depth, which can 
  potentially result in observational signatures if the star's disc is 
  spatially resolved. In particular, this information is present in the star's 
  limb-darkening profile, which is the variation of intensity from the disc centre to 
  the limb.
  Only a few observational methods such as stellar interferometry, analyses of eclipsing
  binaries, transiting 
  extrasolar planets and gravitational microlensing are able to
  constrain stellar limb-darkening in suitable cases.
  Every single measurement thus provides an important 
  opportunity for testing stellar atmosphere models. 

  A Galactic gravitational microlensing event \citep{Paczynski1986} occurs when 
  a foreground massive object passes in the vicinity of the line-of-sight to a 
  background star, resulting in a transient brightening of the source star (called 
  magnification, or amplification). 
  Microlenses can spatially resolve a source star thanks to caustic structures created by a
  lens. They are formed by a single point or by a set of closed curves, along which the
  point-source magnification 
  is formally infinite, with a steep increase in magnification in their vicinity. 
  In practice, this increase is so steep that the characteristic length scale of the differential
  magnification effect is of the order of a fraction of the source star's radius.
  Early works by e.g. \cite{Witt1995} or \cite{Loeb1995} 
  have pointed out the sensitivity of microlensing light curves to limb-darkening,
  with the aim to help remove microlensing model degeneracies. The specific use of
  microlensing as a tool to study stellar atmosphere was proposed later
  \citep[e.g.][]{Valls-Gabaud1995,Sasselov1996,Hendry1998,Gaudi1999}, in particular to 
  probe Galactic bulge red giant atmospheres \citep{Heyrovsky2000}. Indeed, for a given 
  microlensing configuration, the spatial resolution increases with the source's 
  physical diameter, so that giant stars are primary targets.  

  Limb darkening measurements  by microlensing were performed for a number of main-sequence 
  and giant microlensed stars. Event MACHO~1998-SMC-1 \citep{Albrow1999a,Afonso2000} allowed for the 
  first time such a measurement for a metal-poor A6 dwarf located in
  the Small Magellanic Cloud (SMC). 
  Its stellar type was derived from a spectroscopic and photometric analysis in five filters; the
  lens was a binary star also located in the SMC. No real comparison with atmosphere models
  could be provided since very little data existed for these metal-poor A stars.
  The first microlensing limb-darkening measurement for a solar-like star was reported 
  by \citet{Abe2003}: the source was identified as an F8-G2 main-sequence turn-off star,
  involved in the very high-magnification microlensing event MOA 2002-BLG-33
  caused by a binary microlens. A good agreement with limb-darkening coefficient
  predictions was obtained in the $I$ band.
  A limb-darkening measurement for the late G / early K sub-giant was also performed
  by \citet{Albrow2001a} with the binary-lens caustic-crossing event OGLE~1999-BLG-23. The 
  stellar type of the source star was identified by comparing its position
  on two colour-magnitude diagrams obtained from two different telescopes, and deriving
  the star's effective temperature from colour calibration. Again, they found a good
  agreement with stellar models both for the $I$ and $R$ filters.

  Most of the limb-darkening measurements, however, were obtained on Galactic-bulge giant stars. 
  The first case was reported by \citet{Alcock1997} for MACHO~95-30, which involved a very late M4
  red giant source star (spectroscopic typing). 
  In this event theoretical limb-darkening coefficients were only used to improve the light-curve
  fit, but no limb-darkening measurement has been performed. \cite{Heyrovsky2003} later argued that
  the intrinsic variability of the source star precluded any useful limb-darkening analysis.
  Late M giants are of special interest because they give access to testing models at the lower end
  of the temperature range used to compute most of the synthetic model atmosphere grids.  
  For the event  MACHO~1997-BLG-28, \citet{Albrow1999} derived $I$ and
  $V$ coefficients for a K2 giant (typing from
  spectroscopic observations) crossing a caustic cusp, 
  and found a good agreement with stellar models predictions. However, in such a
  complex event, many side effects could have affected the light curve, which somehow decrease the
  strength of the conclusions. Such a remark holds as well for MACHO~1997-BLG-41
  \citep{Albrow2000}, which involved a late G5-8 giant crossing two disjoint caustics. 

  Microlensing event EROS BLG-2000-5 provided the first very good opportunity to test 
  at high precision the limb-darkening of a K3 giant (typing based on both photometry and high-resolution
  spectroscopy) in five filters \citep{Fields2003}. 
  They concluded that their results in the $V$, $I$, and $H$ filters were discrepant 
  from atmosphere models, and furthermore argued 
  that the discrepancy is unlikely to be due to microlensing light-curve modelling 
  drawbacks, but could rather
  be explained by inadequate physics in the stellar models that may be not applicable for all surface
  gravities. A clear variation with time in the shape and equivalent width of the \element{H}$\alpha$ line
  was also reported for the first time in this event \citep{Afonso2001,Castro2001}.
  Limb-darkening was also detected in OGLE~2003-BLG-238 \citep{Jiang2004} and OGLE~2004-BLG-262
  \citep{Yoo2004}, which involved early K1-2 giants, but no strong conclusions on limb darkening
  could be drawn from these events. 

  From the binary-lens event OGLE~2002-BLG-069 \citep{Cassan2004,Kubas2005}, it was possible 
  to obtain not only limb-darkening measurements for a G5 bulge giant source star in the $I$ and $R$ bands,
  but also to directly test predictions from PHOENIX stellar model atmospheres by comparing 
  the change of the \element{H}$\alpha$ equivalent width during a caustic 
  crossing \citep{Cassan2004,Thurl2006} using high-resolution UVES/VLT spectra.
  A discrepancy was found between model and observations, which is most probably explained by the
  lack of a proper chromosphere implementation in the used stellar models.
  More recently, \citet{Cassan2006} performed limb-darkening measurements for the K3 giant source of
  OGLE~2004-BLG-254, and furthermore discussed an apparent systematic discrepancy between stellar
  model predictions and measurements that is observed for G-K bulge giants. However, in the case of 
  OGLE~2004-BLG-254, it appeared that fitting all data sets together or only a subset of them had an
  influence on the limb-darkening measurements \citep{Heyrovsky2008conf}, which remove the observed
  discrepancy. In order to quantify this effect, we provide in this paper a detailed study on
  the impact of including data sets on the resulting limb-darkening measurements. 

  We model and analyse \event, a relatively high-magnification single-lens
  microlensing event that exhibits clear extended-source effects. The source-star fundamental
  parameters and spectral typing were derived from a high-resolution spectrum obtained on VLT/UVES
  as part of a ToO programme. A good multi-site and multi-band coverage of the light curve allows us to
  extract linear limb-darkening coefficients, which we compare to model-atmosphere predictions. 

  The paper is organised as follows: in \Sec{sec:photometry} we
  present the \event event, our photometric
  data and our data reduction procedures. We perform a detailed modelling of the light curve in
  \Sec{sec:models}. The fundamental properties of the target source star are derived in
  \Sec{sec:srceparam}. Section 5 is dedicated to a detailed analysis of the measured
  linear limb-darkening coefficients and their comparison with model-atmosphere predictions. 
  Finally  in \Sec{sec:PCA} we perform a similar analysis using an alternative description of
  limb-darkening based on a principal component analysis of a set of ATLAS limb-darkening profiles.

  \section{Photometric data} \label{sec:photometry}

  \begin{figure*}[!ht]
    \begin{center}
      \includegraphics[width=18cm]{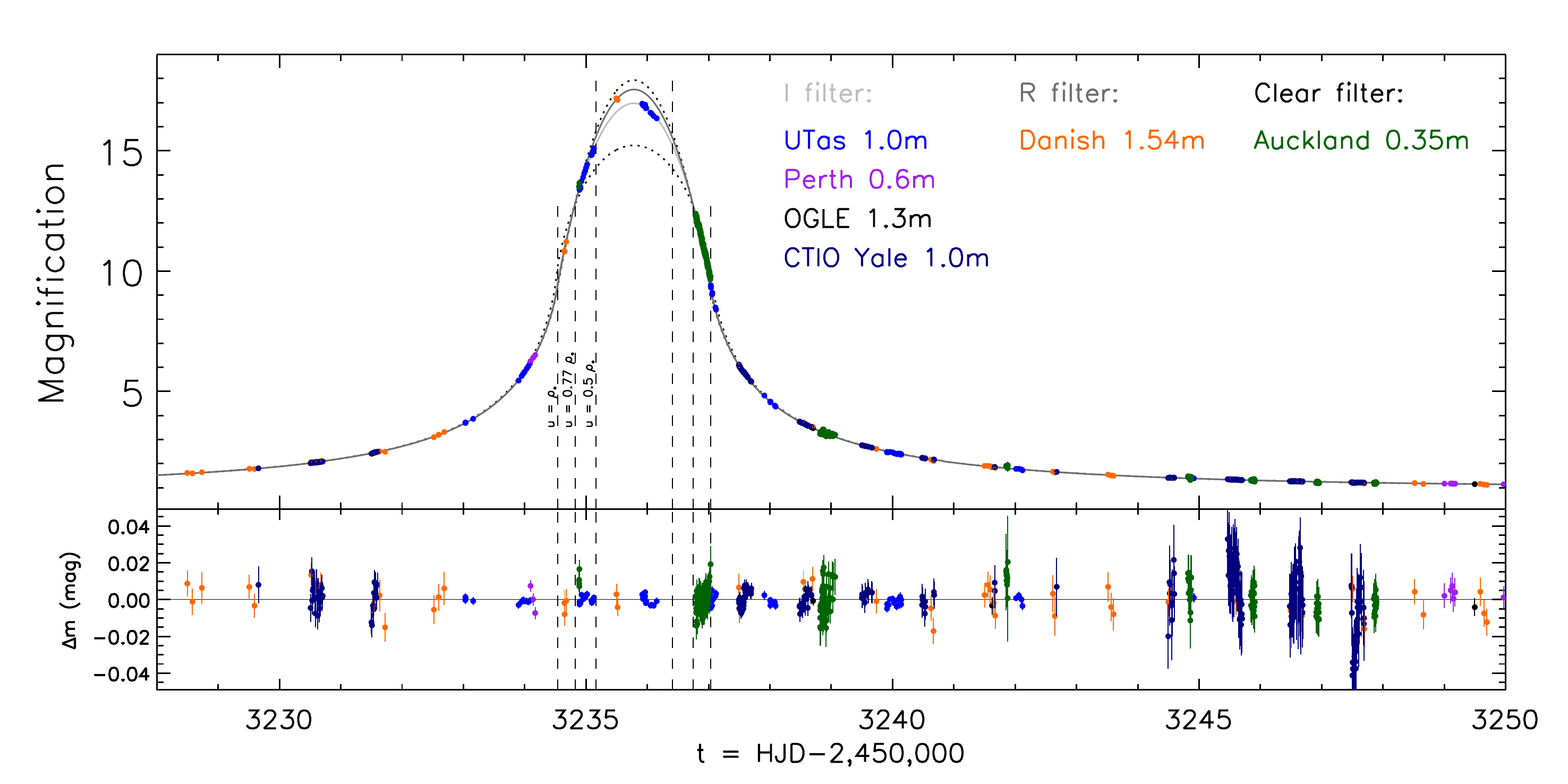}           
      \caption{
	Light curve of \event, with data from PLANET 
	(Danish, UTas and Perth), OGLE and $\mu$FUN 
	(CTIO-Yale and Auckland) collaborations. The two gray solid lines in the upper
	panel draw the best-fit model for the $I$ and $R$ filters with
	linear limb-darkening parameters given in
	\Tab{tab:fits}. The two dotted curves correspond
	to the two extreme cases, $\Gamma = 0$ (uniformly
	bright source, lower dotted curve) and $\Gamma = 1$ (upper dotted
	curve). The two pairs of vertical dashed lines marked $u=\rhoS$ and 
	$u=0.5\,\rhoS$ indicate when the lens is located at
	the limb of the source and half way from its centre to the limb.
	All the curves intersect at $u=0.77\,\rhoS$, also marked by 
	a vertical dashed line. 
	The fit residuals in magnitudes are displayed in the lower panel. } 
      \label{fig:lc}
    \end{center}
  \end{figure*}
   
  \subsection{Event alert and follow-up observations}  \label{sec:data}
  
  The Galactic microlensing event \event ($\alpha=$~17h57m30.6s,
  $\delta=-30\degr 51\arcmin 30\farcs 1$ (J2000.0), or $l=-0.3392\degr$, 
  $b=-3.1968\degr$) was discovered and publicly alerted on August 8, 2004 
  (MHJD\footnote{Modified Heliocentric Julian Date: ${\rm MHJD} = {\rm HJD}-2,450,000$}$\simeq 3226$) by 
  the OGLE-III\footnote{http://ogle.astrouw.edu.pl}
  Early Warning System \citep[``EWS'',][]{Udalski2003} on the basis
  of observations carried out in the $I$ band with the $1.3$~m Warsaw
  Telescope at Las Campanas Observatory (Chile).
 
  \begin{table}[!ht]      
    \begin{center}
      \caption{
	Final selection of data sets, with the raw
	number of observational data (frames) and our final
	selection after the cleaning process.
	The last column lists the adopted error-bar rescaling factors.}
      \begin{tabular}{lccccc}
	\hline
	Telescope  & Filter  & Data & (Frames) & $k_\sigma$ \\
	\hline\hline
	UTas (PLANET)        & I       & 86  & (128) & 2.4  \\
	Perth (PLANET)       & I       & 13  & (15)  & 3.8  \\
	OGLE                 & I       & 44  & (68)  & 2.4  \\
	CTIO-Yale ($\mu$FUN) & I       & 233 & (285) & 4.2  \\
	Danish (PLANET)      & R       & 51  & (67)  & 3.2  \\
	Auckland ($\mu$FUN)  & (clear) & 266 & (334) & 2.4  \\
	\hline
	All data   &  $-$    & 693 & (897) & $-$  \\
	\hline
      \end{tabular}
      \label{tab:datasets}
    \end{center}
  \end{table}

  Following this alert, the PLANET collaboration 
  ({\it Probing Lensing ANomalies NETwork})
  started its photometric 
  follow-up observations on August 10, 2004 (${\rm MHJD} \simeq 3228$), 
  using a network of ground-based telescopes, including the Danish~1.54m 
  (La Silla, Chile), Canopus~1m (Hobart, Tasmania),
  and Perth/Lowell 0.6m (Bickley, Western Australia) telescopes. 
  Data sets and quasi real-time fitted light curves were made
  publicly available online\footnote{http://planet.iap.fr}, as part of a 
  general data sharing policy.
  The event was also monitored by the $\mu$FUN 
  collaboration\footnote{http://www.astronomy.ohio-state.edu/$\sim$microfun}, 
  which gathered data from six
  telescopes: the 1.3m and Yale~1.0m (Cerro Tololo
  Inter-American Observatory, Chile), 
  the Palomar 1.5m telescope (Palomar Observatory, USA), Wise~1m 
  (Mitzpe Ramon, Israel), and two New Zealand amateur telescopes at Auckland (0.35m)
  and Farm Cove (0.25m). 
  
  On August 15, 2004 (${\rm MHJD} \simeq 3233$), photometric data indicated a
  deviation from a normal point-source point-lens light curve. A public
  alert was issued on August 16, 2004 16:05 UT, pointing towards 
  a high peak magnification event, possibly featuring strong extended-source 
  size effects.
  In the following hours, on August 17, 2004
  a target of opportunity (ToO) was activated on the UVES spectrograph at ESO
  VLT in order to monitor the event peak magnification region where
  spectroscopic effects are expected. Thanks to an almost real-time
  modelling operated in parallel, the crossing time of the source disc
  by the lens was estimated to be around $2.4$ days. The peak of the light curve
  was reached  on August  18, 2004 18:32 UT
  at almost three magnitudes above the baseline, corresponding to a minimum
  ({\it i.e.} with null blending) peak magnification of $A \sim 15$.

  \subsection{Data reduction and error bars} \label{sec:data}

  The OGLE data were reduced with their own pipeline, while PLANET and $\mu$FUN data
  were reduced with various versions of the PLANET pipeline \citep[pySIS;][]{pySIS2009}.
  All these reductions are based on the image-subtraction method
  \citep{AlardLupton1998,Bramich2008}.
  A preliminary image-quality inspection 
  helped to remove images with a significant gradient across the
  field, owing to strong background moonlight.
  Under-exposed images were also removed in this process.
  We paid particular attention to the quality of data taken at La Silla 
  at the time of peak magnification, because of unfavourable weather 
  conditions at that site. We could however keep a few trusted data points.
  
  After the data reduction process, we set for each PLANET and $\mu$FUN 
  telescope a range of seeing and background within which the homogeneity 
  of the data sets is ensured. For the Yale telescope, we excluded
  data with seeing outside the range $1.8$--$3.2"$.
  In the case of UTas data, we applied an upper limit on the seeing of
  $3.0"$, and for the Perth, Danish, and Auckland telescopes, $3.3"$.
  Our final data set is presented in \Tab{tab:datasets} and displayed
  in \Fig{fig:lc} (Auckland telescope had no filter at the time of the observation,
  so the filter is referred to as \textit{clear}).
  
  It is known that the error bars we obtain from the data reduction are usually
  underestimated, and are not homogeneous from one data set to another. To avoid
  this problem, we rescaled the error bars, so that from the best model one has 
  $\chi^2/N \simeq 1$ for each data set fitted alone, with $N$ the
  corresponding number of data points. 
    Moreover, it happens that some of the original error bars $\sigma$
    are unrealistically small; to prevent this, we added in quadrature an additional
    term to the rescaled error bars, so
    that ${\sigma^{\prime}}^2=(k_{\sigma}\,\sigma)^2+(4\dix{-4})^2$ magnitudes. 
    When the original error bars are small (UTas and Danish), the constant term 
    dominates the error bars' size.
    The values of $k_{\sigma}$ are given in \Tab{tab:datasets}.

 \section{Light-curve modelling} \label{sec:models}

 \subsection{Linear limb-darkening formalism} \label{sec:linearLD}
 
 Limb-darkening profiles of stars can be described analytically at different levels of approximation, in
 particular by a sum containing powers of $\mu=\cos\alpha$, where $\alpha$ is the angle
 of a given emerging light ray with respect to the normal of the stellar surface
 \citep[e.g.][]{Claret2000}. 
 In the first degree of approximation, called the \textit{linear limb-darkening} (hereafter, LLD)
 law, the star 
 brightness profile can be written as
 \begin{equation}
   I(r) = 1 - a\,\left( 1-\sqrt{1-r^2}\right)\, ,
   \label{eq:ldatm}
 \end{equation}
 where $r=\sqrt{1-\mu^2}$ is the fractional radius on the stellar disc from where the light is emitted,
 and $a$ is the \textit{linear limb-darkening coefficient} (hereafter LLDC). 
 In this work, we will concentrate on measuring LLDCs. Firstly, because in microlensing events higher
 order coefficients have a very small impact, e.g. \cite{Dominik2004ld} finds that for a caustic
 crossing, the 
 effect of the change of the LLDC on light curve is $\sim 25$ times greater than the square-root  
 coefficient. Secondly, because a strong correlation exists between the
 coefficients, it is impossible to precisely measure the LLDC when a further coefficient is taken
 into account \citep{Kubas2005}. Lastly, because LLDC are widely used and are available
 in catalogues; it is an important aspect for our goal to compare our results 
 with the existing literature.

 For our modelling purpose, a more convenient way to rewrite the LLD law is
 to have a formula that conserves the total source flux for all LLDC values. With this
 requirement, the LLD law equivalent to \Eq{eq:ldatm} \citep {Albrow1999c} 
 but normalised to unit flux can be written as
 \begin{equation} 
   I(r) = \frac{1}{\pi} \left[ 1 - \Gamma \left(1-\frac{3}{2}\
     \sqrt{1-r^2}\right) \right]\, , 
   \label{eq:ldnorm}
 \end{equation}
 where $\Gamma$ is the LLDC modelling parameter, with
 \begin{equation} 
   a = \frac{3\,\Gamma}{2+\Gamma}\, .
   \label{eq:ldgam}
 \end{equation}
 With this formalism, it is interesting to notice that all limb-darkening profiles
 intersect at a common fractional radius \citep {Heyrovsky2003}, 
 $r_{\rm lld} = \sqrt{5}/3 \simeq 0.75$.

 \subsection{Single-lens, extended-source models}

 In its motion relative to the lens, the source centre approaches the lens at a minimal distance
 $\uo$ in units of the angular Einstein ring radius $\ThE=\sqrt{4GMc^{-2}\,(\DL^{-1}-\DS^{-1})}$
 \citep[][with $\DS$, $\DL$ the distances from the source and the lens to the observer, $M$ the lens
 mass]{Einstein1936}, which can be smaller than the source radius $\rhoS$ expressed in the same units.
 Because high-magnification events involve low values of the impact parameter $\uo$,
 they are likely to be affected by extended-source effects in particular if the
 source star is a giant. Although this happens fairly rarely in practice (a couple of cases amongst
 the $\sim 700$ microlensing events observed every year), this is the case for \event.

 The point-source magnification at the exact location of the lens is formally infinite,
 following the well-known formula \citep{Paczynski1986}
 \begin{equation} 
   \Apspl(u) = \frac{u^2+2}{u\sqrt{u^2+4}}\, ,
   \label{eq:pspl}
 \end{equation}
 where $u$ is the distance from the lens to a given point on the source in units of $\ThE$. 
 Consequently, the flux originating from regions of the source in the immediate
 neighbourhood of the lens (typically a fraction of the source radius) is preferentially
 amplified. The relative motion of the source and lens then results in a time-dependent probing
 of the stellar atmosphere at different fractional radius, corresponding to different optical depths.

 Single-lens light curves affected by extended-source effects display a characteristic
 flattening at their peak. For a uniformly bright extended source, 
 \cite{WittMao1994} derived an exact analytic formula for the magnification, which involves
 elliptic integrals. But there is no similar formula to describe limb-darkened sources, and
 calculating the exact magnification requires numerical integration. One way is to decompose the
 source into small rings of uniform intensity. Another approach by \cite{Heyrovsky2003} is to perform
 the angular integration over the stellar disc analytically and only the radial integration
 numerically, for arbitrary sources.

 If some conditions are fulfilled, it is also possible to use approximate formulae, which have the
 advantage to allow us a very fast computation. Considering that in \Eq{eq:pspl}, $\Apspl \simeq 1/u$ when
 $u \ll 1$, \cite{Yoo2004} find that the magnification $\Aext$ for an extended source with a
 linear limb-darkening profile with coefficient $\Gamma$ can be expressed as
 \begin{eqnarray}
   & & \Aext\left( u, \rhoS \right) = \left[ B_0\left( z \right) -
     \Gamma\, B_1\left( z \right) \right] \, \Apspl\left( u
   \right) \, , \nonumber\\ 
   & & z = u/\rhoS \, , \nonumber\\
   & & B_0\left( z \right) = \frac{4\,z}{\pi} \, E\left[ \arcsin \min \left(
       1, \frac{1}{z} \right), z \right] \, ,\\
   & & B_1\left( z \right) = B_0\left( z \right) - \frac{3\,z}{\pi}
   \,\int\limits_{0}^{\pi} \int\limits_{0}^{1} \frac{r \sqrt{1-r^2} 
   }{\sqrt{r^2+z^2-2\,z\,r \cos \phi}} \,\d r \,\d\,\phi\, ,
   \nonumber
   \label{eq:Yoo}
 \end{eqnarray}
 where $E$ is the incomplete elliptic integral of the second kind following the 
 notation of \citet{GradshteynRyz1965}. The integral $B_1$ can be
 efficiently evaluated and tabulated for $z$, as can $B_0$. This approximation is valid as
 far as $ \rhoS^2/8 \ll 1$ and $\uo \ll 1$. Because these
 relations hold for \event we choose this formalism (although close to
 the limit case of application, since the maximum error for a uniform source here is of the order of
 $0.2$\%, but is still much lower than the photometric errors).

 The complete model then involves four parameters: the source radius $\rhoS$, as well as
 $\uo$, $\to$ and $\tE$, which define the
 rectilinear motion of the source with respect to the lens, so that the lens-source separation $u$
 satisfies $u^2(t)=\uo^2+(t-\to)^2/\tE^2$. Moreover, one has to take
 into account for each telescope ``$i$'' two more parameters, the baseline magnitude
 $M_b^i=-2.5\,\log(\FS^i+\FB^i)$ and the blending factor $g^{i}=\FB^{i}/\FS^{i}$. Here, $\FS^i$ 
 and $\FB^i$ are the source and the blend flux, the latter referring to any un-magnified
 flux entering the photometric aperture, from the lens itself and e.g. background stars.
 They are related to the time-dependent magnification 
 $\Aext$ by $F^i(t)= \Aext(t)\,\FS^i+\FB^i$.

 \subsection{Fitting procedure} \label{sec:fitting}

 To fit our data sets, we use two minimisation schemes: Powell's method
 and a Markov-Chain Monte-Carlo algorithm, from which we also obtain the model
 parameter error bars \citep{Kains472,UCAUSFIX}. As stated before, it is impossible to define a proper number
 of degrees of freedom. Indeed, the parameters $\uo$, $\to$, $\tE$ and $\rhoS$ are common 
 to all data sets, whereas $M_b^i$ and $g^i$ are associated to the data set ``$i$'', and the LLDCs 
 may be chosen to be common per observing filter or per individual telescope. This explains
 the choice of $N$ instead of $\rm{d.o.f}$ to rescale the error bars in \Sec{sec:data}.
 The first requirement to get precise measurements of limb-darkening coefficients is to get
 an overall well-covered light curve. This allows us to secure good measurements of the basic parameters 
 $\uo$, $\to$, $\tE$ and $\rhoS$, as well as $M_b^i$ and $g^i$. The region of the light curve with
 a noteworthy sensitivity to limb-darkening is, however, mainly limited to when the lens is inside the
 source-star disc, and drops to a few percent outside \citep{Heyrovsky2000}. 
 We now discuss this aspect in greater detail.
 
 While all limb-darkening profiles intersect at the same fractional radius 
 $r_{\rm lld} \simeq 0.75$ as seen in \Sec{sec:linearLD}, the corresponding
 magnification light curves intersect at around $u_{\rm lld} \simeq 0.77\,\rhoS$ 
 (with $u$ the lens-source centre distance). This special point is marked by 
 a vertical dashed line in \Fig{fig:lc}, in which we have also indicated two other 
 interesting positions of the lens: at the limb of the source ($u=\rhoS$) and
 at half-way from its centre to its limb ($u=0.5\,\rhoS$). The two dotted magnification
 curves of the figure show the two extreme cases of LLDC, $\Gamma=0$ (no limb-darkening) 
 and $\Gamma=1$. From this we can distinguish three main regions: $0<u/\rhoS<0.5$, where the
 limb-darkening sensitivity is high, up to $\sim 16$\% ;
 $0.5<u/\rhoS<0.77$, where the sensitivity decreases outward to $0$, 
 and $0.77<u/\rhoS<1$ where
 the sensitivity increases outward and peaks at the limb at $\sim$ 8\% 
 \citep{Heyrovsky2003}.
 Based on this argument and from our data coverage of \event shown in \Fig{fig:lc},
 it is clear that we can expect LLDC measurements from UTas $I$-band, Danish $R$-band and
 Auckland's clear-filter.

 The best-fit parameters and their error bars are given in \Tab{tab:fits} for different
 combinations of data sets. We comment on the results in detail in Sects
 \ref{sec:LDmesure} and
 \ref{sec:PCA}. In \Fig{fig:lc} we plot the combined fit including all telescopes and using one
 coefficient per band.

 \subsection{Estimates of the lens properties} \label{sec:const}

 Although the properties of the lens are not of primary interest here,
 we can still provide an estimate of the lens' mass and distance. However, these quantities
 cannot be measured here, because an additional observable, such as the
 parallax, is needed to remove a degeneracy between these two parameters. Here, 
 parallax effects are not visible because the time scale of the event is very 
 short, $\tE \simeq 10$~days~$ \ll 1$~year. 

   From our modelling and our estimate of the source
   radius and distance (see 
   \Sec{sec:CMDcalib}), we derive the Einstein radius to be around 
   $\ThE = \ThS/\rhoS \simeq 0.4$~mas, which leads to a relative
   proper motion of $\mu = \ThE/\tE \simeq 16$~mas/yr.
   This high proper motion almost certainly means that the
   lens is located in the disc (or possibly in the halo). Moreover, with such
   a high motion, there is a good chance that the lens can be
   clearly visible (away from the source) in a few years by adaptive
   optics observations. 

   \begin{figure}[!ht]
   \begin{center}
     \includegraphics[width=8.7cm]{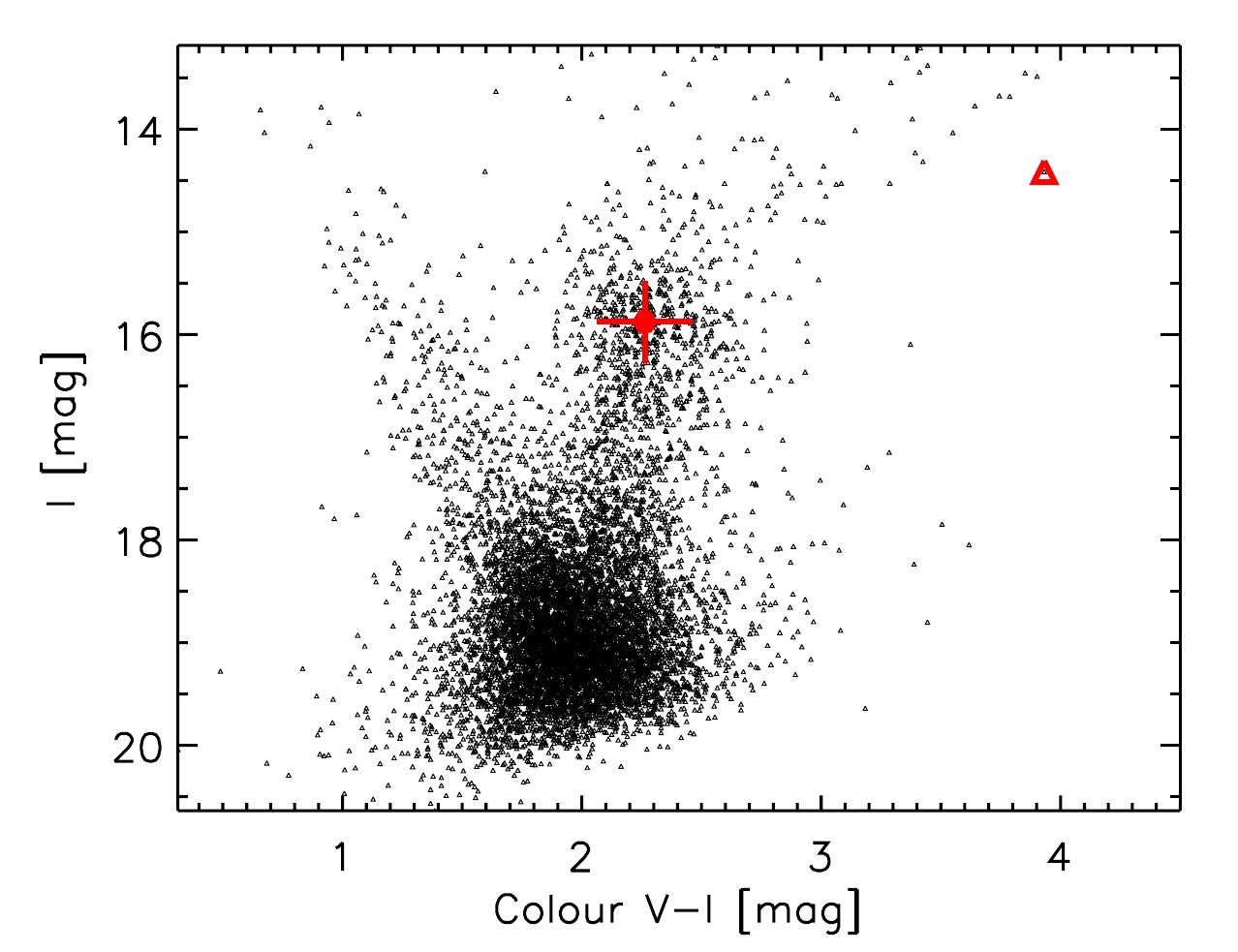}        
     \caption{ 
       OGLE-III BLG182.8 field calibrated $I$ vs. $(V-I)$ colour-magnitude diagram,
       comprising stars within a radius of $2.16\arcmin$ centred on our target 
       OGLE~2004--BLG--482 (red open triangle). The red circle indicates 
       the mean position of the RCG centre, and the cross the width of
       the two-dimensional Gaussian distribution.}      
     \label{fig:cmd}
   \end{center}
 \end{figure}

 \section{Source star properties} \label{sec:srceparam}

 \subsection{OGLE calibrated colour-magnitude diagram} \label{sec:CMDcalib} 

 The microlensing event OGLE 2004-BLG-482 
 occurred in OGLE-III BLG182.8 field,
 and was also observed during the second phase of OGLE in field BUL\_SC23.
 From the calibrated photometry in $I$ and $V$ filters of the OGLE-III BLG182.8
 field, we extract an $I$ vs. $(V-I)$ 
 colour-magnitude diagram (CMD) by selecting stars surrounding our 
 target within a circle of radius $2.16\arcmin$ ($\sim 9000$ stars),
 as shown in \Fig{fig:cmd}.
 This choice ensures that we have enough stars 
 to construct the CMD while keeping a reasonably homogeneous extinction 
 across the selected region. 

 Our target is indicated as the red open triangle
 and has a calibrated magnitude and colour of $I = 14.41 \pm 0.03$ and
 $(V-I) = 3.93 \pm 0.04$. 
  From the analysis of the OGLE images, we conclude that this bright target is 
  not blended by a neighbouring star in $I$. We also checked in the OGLE-III 
  photometric catalogue, which has a better resolution than OGLE-II, that there
  is no blended star within 1 arcsec bright enough in $V$ to contaminate our
  measurements. Finally, our model finds a blending ratio close to zero, 
  justifying our assumption that the measured magnitude and colour of
  the target can safely be assigned to the source star.

  In order to correct these measurements for extinction, we can either use the
  reddening maps of \citet{Sumi2004} based on OGLE-II photometry, or use the red
  clump giant (RCG hereafter) assuming that our source suffers the same amount 
  of extinction. At the position of the source, \citet{Sumi2004} measures an 
  extinction of $\EVI = 1.405 \pm 0.027$. This is derived assuming a clump
  colour of 1.028 and a ratio of total to selective extinction $\Ri = 0.964$,
  giving an absorption of $\Ai = \Ri\times\EVI = 1.36 \pm 0.06$.
 
 \begin{figure*}[!ht]
   \begin{center}
      \includegraphics[width=14.2cm,angle=90]{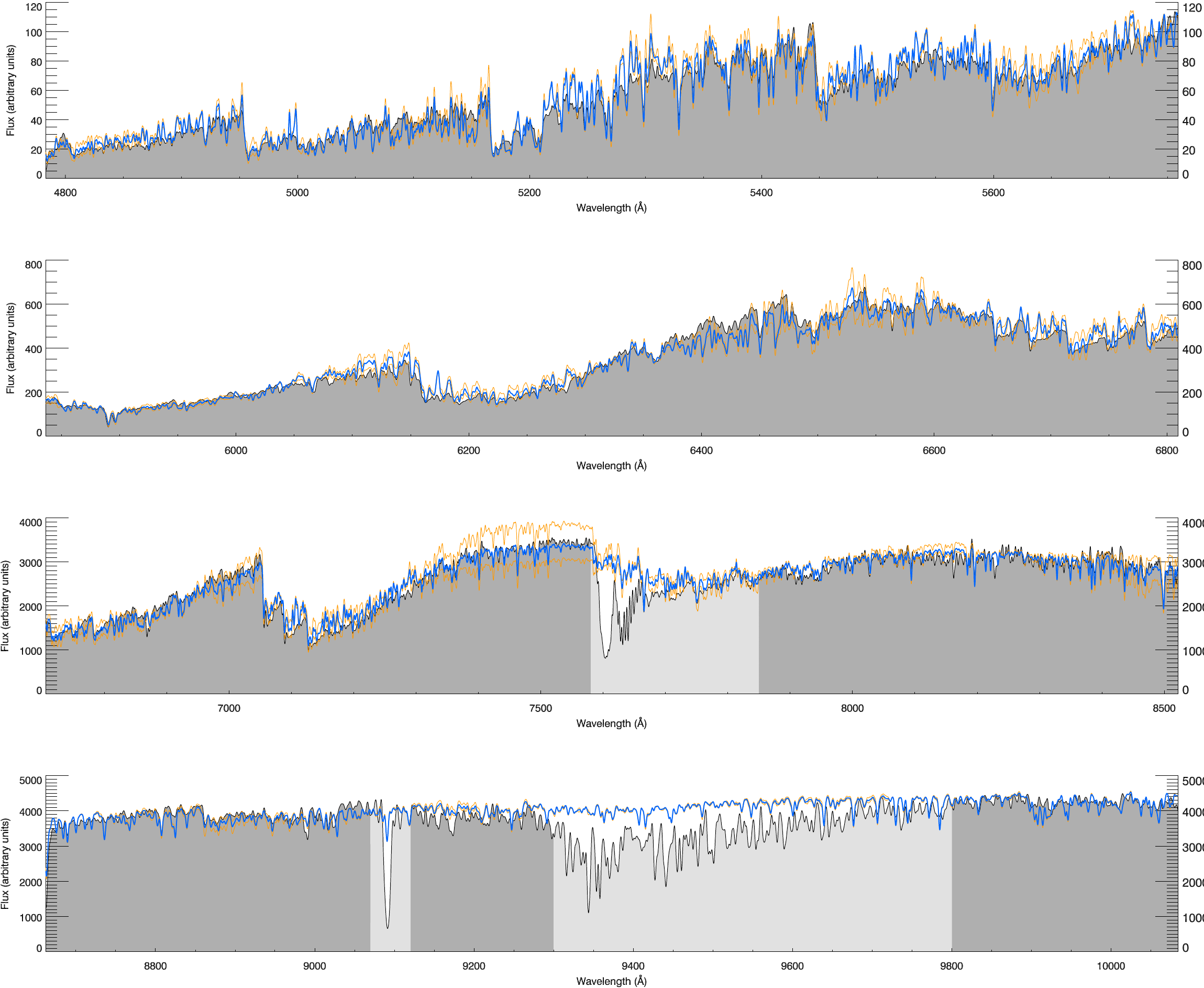}    
     \caption{
       Observed (black line) and best-fit (blue) template spectrum of \event. 
       The region around the TiO 7100 shows the agreement of the observed and synthetic 
       spectra. The two orange curves are plotted at $\pm 100$~K.
       The regions excluded from the fitting process are shaded in
       light grey,
       while the remainder of the spectrum is shaded in dark grey. }
     \label{fig:spectra}
   \end{center}
 \end{figure*}

 The RCG central position is marked in \Fig{fig:cmd}
 as a red circle with error bars. To determine its mean magnitude and colour, 
 we fit a two-dimensional Gaussian around its position
 ($\sim 400$ stars), from which we derive $I_{\rm RCG} = 15.88 \pm 0.01$ and  
 $(V-I)_{\rm RCG} = 2.263 \pm 0.004$. 
  Using the same clump colour as \citet{Sumi2004}, we get an extinction of
  $\EVI = 1.235$. Given the uncertainty of the intrinsic clump colour, due to
  variations with age and metallicity \citep{Salaris2002}, this estimate agrees
  with the previous one. We therefore adopt as the dereddened magnitudes and 
  colour of our target $\Io = 13.05 \pm 0.07$, $\VIo = 2.53 \pm 0.05$ and
  $\Vo = 15.58 \pm 0.09$.

  In principle the observed position of the clump could be used to measure its
  distance. In practice, the absolute magnitude of the clump depends on age and
  metallicity, and corrections introduce an uncertainty as for its intrinsic
  colour. Moreover, the reddening corrections are not accurately determined. We
  therefore prefer fixing the distance and check that the observed clump 
  position is compatible with this choice. We start by assuming a Galactic
  Centre distance of $8.33 \pm 0.35$ kpc from \citet{Gillessen2009}, giving a
  distance modulus of $14.60 \pm 0.09$. Then, we assume that Baade's Window at
  ($l=1.00^{\circ}$ $b=-3.88^{\circ}$) is at about the same distance as the 
  Galactic Centre, according to \citet{Paczynski1998}. This fixes the distance 
  to OGLE-II field BUL\_SC45, which contains Baade's Window. Finally, we use 
  the relative distance of field BUL\_SC23 with respect to BUL\_SC45 as given
  by \citet{Rattenbury2007} (their Table 1), which amounts to 0.13 mag, to get
  a distance modulus of $14.73 \pm 0.15$ to the clump in the direction of our 
  target. The corresponding absolute magnitude of the clump is then
  $15.88 - 1.36 - 14.73 = -0.21$, in good agreement with the most recent value
  determined by \citet{Groenewegen2008} for the local red clump based on 
  revised Hipparcos data, namely $\MiRC = -0.22 \pm 0.03$.

  If our source is at the same distance as the clump in its direction and 
  suffers the same amount of extinction, its expected dereddened magnitude is
  $14.41 - 15.88 + 14.51 = 13.04$, in good agreement with the previous estimate
  based on Sumi's reddening law. The agreement in colour is not as good, at
  $3.93 - 2.263 + 1.028 = 2.70$.
 
  We then fit calibrated isochrones from \cite{Bonatto2004} to 2MASS
  data in
  our field, to derive the following near-infrared extinctions: 
  $\AJ = 0.52 \pm 0.10$, $\AH = 0.36 \pm 0.11$ and $\Aks = 0.20 \pm 0.02$. 
  From this and the magnitudes listed in the 2MASS PSC for our target
  (2MASS 17573061-3051305), we get $J_{0} = 11.55 \pm 0.10$, 
  $H_{0} = 10.68 \pm 0.11$ and $\Kso = 10.42 \pm 0.04$, and the corresponding 
  colours $\JHo = 0.87 \pm 0.16$, $\HKo = 0.26 \pm 0.12$ and 
  $\JKo = 1.13 \pm 0.11$. Converting to Bessell \& Brett near-infrared 
  photometric system \citep{Bessell1988} gives $\JKo = 1.17$ and $\Ko = 10.46$,
  corresponding to an M4 giant (their Table 3), which have mean colours of 
  $\VIo = 2.55$ and $\VKo = 5.10$, in good agreement with our observed values
  $\VIo = 2.53$ and $\VKo = 15.58 - 10.46 = 5.12$.

  This allows us to estimate the source radius using the surface brightness 
  relation: $\log\ThS+\Ko/5 = (0.037 \pm 0.007)\times\VKo + (0.610 \pm 0.028)$
  from \cite{Groenewegen2004} calibrated on 40 M giants, where $\ThS$ is the 
  source angular diameter in $mas$. We find an angular diameter of 
  $\ThS = 51 \pm 3 \ \muas$, which at the adopted source distance of 
  $d = 8.8 \pm 0.6$~kpc corresponds to a physical source radius of 
  $\RS = 48 \pm 4 \ \rsun$.

 In the next section, we perform the 
 analysis of the VLT/UVES high-resolution spectra that we obtained on this event, 
 in order to derive more accurately the spectral type and to determine
 the fundamental parameters of the source star.

 \subsection{VLT/UVES spectroscopy}  \label{sec:fundamentalp}

 We have obtained for \event high-resolution
 spectra \textbf{($R\sim 40000$)} on VLT/UVES, as part of a ToO activated shortly after the peak of the 
 light curve was passed. The data were reduced in a standard way using version 2.1 
 of the UVES context of the MIDAS data reduction software.

 The spectrum is dominated by broad absorption bands from
 molecules. The shape and depth of molecular absorption bands, particularly TiO,
 are very sensitive to the stellar effective temperature $\teff$, and to a lesser
 degree also to the surface gravity $\logg$. We estimated the atmospheric
 parameters of \event by comparing the observed spectrum with a
 grid of pre-calculated synthetic template spectra.

 The grid of synthetic template spectra, calculated by Plez (priv. comm.), is
 based on synthetic spectra calculated from MARCS spherical model atmospheres with 1D emergent
 spectra and LTE
 radiative transfer \citep{Gustafsson2008,Gustafsson2003,Gustafsson1975,Plez2003,Plez1992},
 and includes the latest available atomic and molecular  
 line data \citep{Gustafsson2008,Kupka1999,Plez1998}.
 Synthetic template spectra for M giants calculated with the MARCS model
 atmospheres have a good record for determining stellar parameters in M
 supergiants \citep[e.g.][]{Levesque2005,Levesque2007,Massey2008} and 
 were extensively used to
 calibrate M giant photometry \citep{Bessell1998}.  
 
 The grid used in our analysis covers an effective temperature range of $3000
 {\rm K} < T_{\rm eff} < 4000 {\rm K}$, with steps of $100$ K, and a surface gravity
 range of $0.0 < \logg < 3.0$, with steps of 0.5. This grid was calculated for
 giants with solar abundances and no carbon enrichment. Since our grid does not
 cover a range of metallicities, we therefore have no leverage on this parameter.
 We also prepared routines to calculate linear interpolations between the spectra
 in our grid for any given value of $\teff$ and $\logg$. 
 
 We then compared the observed spectrum of \event with template
 spectra across the available range of $\teff$ and $\logg$ and determined
 the goodness-of-fit using the $\chi^2$ diagnostic. In calculating $\chi^2$, we
 used the entire observed spectrum, from approximately 4800 to 10000$\,$\AA, only
 excluding three regions that are strongly affected by telluric absorption
 (7580--7850, 9070--9120$\,$\AA  and 9300-9800$\,$\AA). However, since no continuum is
 present in our spectrum, and we also do not know the absolute stellar flux, we
 renormalised the synthetic spectrum using a one-dimensional polynomial function
 prior to calculating $\chi^2$. This renormalisation does not affect the shape
 of the broad molecular bands that are important for determining $\teff$
 and $\logg$.
 
 We refined the best values of $\teff$ and $\logg$ using
 parabolic minimisation between the grid points that yielded the lowest value of
 the $\chi^2$ diagnostic. 
 In \Fig{fig:spectra} we illustrate the agreement
 between the observed and best-fit template spectrum, including estimated
 parameter uncertainties, around the highly temperature-sensitive TiO band near
 7100$\,$\AA.
 We find that the parameters that best fit our observed
 spectrum are $\teff = 3667 \pm 150\,{\rm K}$ and $\logg = 2.1 \pm 1.0$, assuming
 solar abundances.
 The quoted error bars are dominated by systematic uncertainties in the synthetic
 spectra and data reduction procedures used, such as flux calibration. Our
 uncertainties are further increased because our grid of template
 spectra was calculated for only one metallicity. 
 The range of effective temperatures we find is compatible with a star
 of MK spectral type between M1 and M5, with the best-fit value giving
 a red giant star a bit later than M3 \citep{Houdashelt2000,Strassmeier2000}. 

 The large error bar on the surface gravity confirms that our spectrum has little 
 to offer in gravity-sensitive diagnostics. However, we can obtain independent 
 constraints on $\logg$: given that the mass of an M giant of 
 1 or 10 Gyr is smaller than $2.3$ and $1\,\msun$
 respectively, using $\logg = \logg_\odot + \log M - 2\times\log\RS$, we find
 the corresponding upper limits of the surface gravity: $\logg = 1.5 \pm 0.2$ 
 and $\logg = 1.1 \pm 0.2$ respectively, taking into account the uncertainty of the 
 source radius. This agrees with our spectroscopic analysis, although favouring 
 the lower boundary.

 \subsection{Conclusion on the source MK type and parameters}  \label{sec:typingconlcusion}
 
 We finally find a good agreement between our photometric and spectroscopic study, with a source star 
 of MK spectral type a bit later than M3. We therefore adopt the fundamental parameters from the
 spectroscopic analysis ($\teff = 3667 \pm 150\,{\rm K}$, $\logg = 2.1 \pm 1.0$, solar metallicity)
 to make our selection of atmosphere models used to compare our limb-darkening measurements to 
 model predictions, as discussed in the next section.
 
 \begin{table*}[!ht]
   \begin{center}
    \caption{
       Model parameters and error bars for different relevant combinations of data sets. The
       measured linear limb-darkening coefficients are indicated in bold face. 
       The data sets are referred to by letters, following
       the convention indicated in the first line of the table.
       The uncertainties on the parameters are indicated in parenthesis
       and apply to the last significant digit.
       Models for which no stable fit or very unrealistic
       results are obtained are marked with the symbol ``$\badfit$'' following the measured value. }
     \begin{tabular}{l|p{2.2cm}p{2.2cm}p{2.2cm}p{2.2cm}p{2.2cm}p{2.2cm}}
       \hline
       \multicolumn{1}{l|}{Parameters} & \textsc{UTas (U)} & \textsc{Danish (D)} & \textsc{Auckland (A)} &
       \textsc{OGLE (O)} & \textsc{Perth (P)} & \textsc{CTIO Yale (C)} \\
       \hline\hline
       \multicolumn{7}{c}{\bf\textit{Independent fits for \textsc{U}, \textsc{D} and \textsc{A}}} \\
       \hline
       $\to$ (days) & $3235.78(4 \pm 1)$ & $3235.78(3 \pm 3)$ & $3235.76(8 \pm 4)$ & -- & -- & --  \\
       $\uo$ & $0.010(8 \pm 4)$ & $0.0(2 \pm 1)$ & $0.0(0 \pm 1)$ & -- & -- & --  \\
       $\tE$ (days) & $8.(9 \pm 1)$ & $9.(6 \pm 4)$ & $9.(3 \pm 3)$ & -- & -- & --  \\
       $\rhoS$ & $0.14(0 \pm 2)$ & $0.1(3 \pm 1)$ & $0.14(0 \pm 7)$ & -- & -- & --  \\
       {\bf\textit{a}} & $\bf 0.677 \pm 0.013$ & $\bf 0.67 \pm 0.22\badfit$ & $\bf 0.76 \pm 0.13$ &-- & -- & --  \\
       $M_b$ & $ 11.5 $ & $ 11.4 $ & $ 13.5 $ & -- & -- & --  \\
       $g$ & $ 7.0 $ & $ 1.4 $ & $ 4.5 $ & -- & -- & --  \\
       $\chi^2$ & $ 82.5 $ & $ 43.2 $ & $ 234.9 $ & -- & -- & --  \\
       \hline\hline
       \multicolumn{7}{c}{\bf\textit{Combined fit including \textsc{U+D}}} \\
       \hline
       $\to$ (days) & \multicolumn{6}{l}{$3235.784(5 \pm 8)$} \\
       $\uo$ & \multicolumn{6}{l}{$0.00(9 \pm 2)$} \\
       $\tE$ (days) & \multicolumn{6}{l}{$9.1(5 \pm 9)$} \\
       $\rhoS$ & \multicolumn{6}{l}{$0.13(7 \pm 1)$} \\
       {\bf\textit{a}} & $\bf 0.674 \pm 0.012$ & $\bf 0.837 \pm 0.018$ & -- &-- & -- & --  \\
       $M_b$ & $ 11.5 $ & $ 11.4 $ & -- & -- & -- & --  \\
       $g$ & $ 7.2 $ & $ 1.3 $ & -- & -- & -- & --  \\
       $\chi^2$ & $ 85.1 $ & $ 57.8 $ & -- & -- & -- & --  \\
       \hline\hline
       \multicolumn{7}{c}{\bf\textit{Combined fit including \textsc{U+A}}} \\
       \hline
       $\to$ (days) & \multicolumn{6}{l}{$3235.780(8 \pm 8)$} \\
       $\uo$ & \multicolumn{6}{l}{$0.00(0 \pm 3)$} \\
       $\tE$ (days) & \multicolumn{6}{l}{$9.(1 \pm 1)$} \\
       $\rhoS$ & \multicolumn{6}{l}{$0.13(8 \pm 2)$} \\
       {\bf\textit{a}} & $\bf 0.714 \pm 0.013$ & -- & $\bf 0.660 \pm 0.023$ &-- & -- & --  \\
       $M_b$ & $ 11.5 $ & -- & $ 13.5 $ & -- & -- & --  \\
       $g$ & $ 7.2 $ & -- & $ 4.5 $ & -- & -- & --  \\
       $\chi^2$ & $ 101.9 $ & -- & $ 286.3 $ & -- & -- & --  \\
       \hline\hline
       \multicolumn{7}{c}{\bf\textit{Combined fit including \textsc{D+A}}} \\
       \hline
       $\to$ (days) & \multicolumn{6}{l}{$3235.77(5 \pm 3)$} \\
       $\uo$ & \multicolumn{6}{l}{$0.00(0 \pm 7)$} \\
       $\tE$ (days) & \multicolumn{6}{l}{$9.(7 \pm 2)$} \\
       $\rhoS$ & \multicolumn{6}{l}{$0.13(4 \pm 8)$} \\
       {\bf\textit{a}} & -- & $\bf 1.0 \pm 0.23\badfit$ & $\bf 0.93 \pm 0.29\badfit$ &-- & -- & --  \\
       $M_b$ & -- & $ 11.4 $ & $ 13.5 $ & -- & -- & --  \\
       $g$ & -- & $ 1.4 $ & $ 4.8 $ & -- & -- & --  \\
       $\chi^2$ & -- & $ 50.6 $ & $ 241.5 $ & -- & -- & --  \\
       \hline\hline
       \multicolumn{7}{c}{\bf\textit{Combined fit including \textsc{U+D+A}}} \\
       \hline
       $\to$ (days) & \multicolumn{6}{l}{$3235.781(4 \pm 9)$} \\
       $\uo$ & \multicolumn{6}{l}{$0.00(0 \pm 4)$} \\
       $\tE$ (days) & \multicolumn{6}{l}{$9.2(9 \pm 6)$} \\
       $\rhoS$ & \multicolumn{6}{l}{$0.13(6 \pm 1)$} \\
       {\bf\textit{a}} & $\bf 0.713 \pm 0.012$ & $\bf 0.881 \pm 0.010$ & $\bf 0.660 \pm 0.011$ &-- & -- & --  \\
       $M_b$ & $ 11.5 $ & $ 11.4 $ & $ 13.5 $ & -- & -- & --  \\
       $g$ & $ 7.3 $ & $ 1.3 $ & $ 4.6 $ & -- & -- & --  \\
       $\chi^2$ & $ 102.7 $ & $ 58.1 $ & $ 287.8 $ & -- & -- & --  \\
       \hline\hline
       \multicolumn{7}{c}{\bf\textit{Combined fit including all telescopes (one LLDC per band)}} \\
       \hline
       $\to$ (days) & \multicolumn{6}{l}{$3235.781(6 \pm 7)$} \\
       $\uo$ & \multicolumn{6}{l}{$0.00(0 \pm 2)$} \\
       $\tE$ (days) & \multicolumn{6}{l}{$9.6(1 \pm 2)$} \\
       $\rhoS$ & \multicolumn{6}{l}{$0.130(9 \pm 5)$} \\
       {\bf\textit{a}} & $\bf 0.714 \pm 0.010$ & $\bf 0.884 \pm 0.021$ & $\bf 0.652 \pm 0.016$ & $\bf 0.714 \pm 0.010$ & $\bf 0.714 \pm 0.010$ & $\bf 0.714 \pm 0.010$  \\
       $M_b$ & $ 11.5 $ & $ 11.4 $ & $ 13.5 $ & $ 14.1 $ & $ 12.7 $ & $ 14.0 $  \\
       $g$ & $ 7.6 $ & $ 1.4 $ & $ 4.8 $ & $ 0.0 $ & $ 0.7 $ & $ -0.8 $  \\
       $\chi^2$ & $ 122.7 $ & $ 51.0 $ & $ 286.6 $ & $ 42.6 $ & $ 14.3 $ & $ 239.7 $  \\
       \hline
     \end{tabular}
     \label{tab:fits}
  \end{center}
 \end{table*}
 
\begin{figure*}[!ht]
   \begin{center}
     \includegraphics[width=5.9cm, bb= 0 0 227 567]{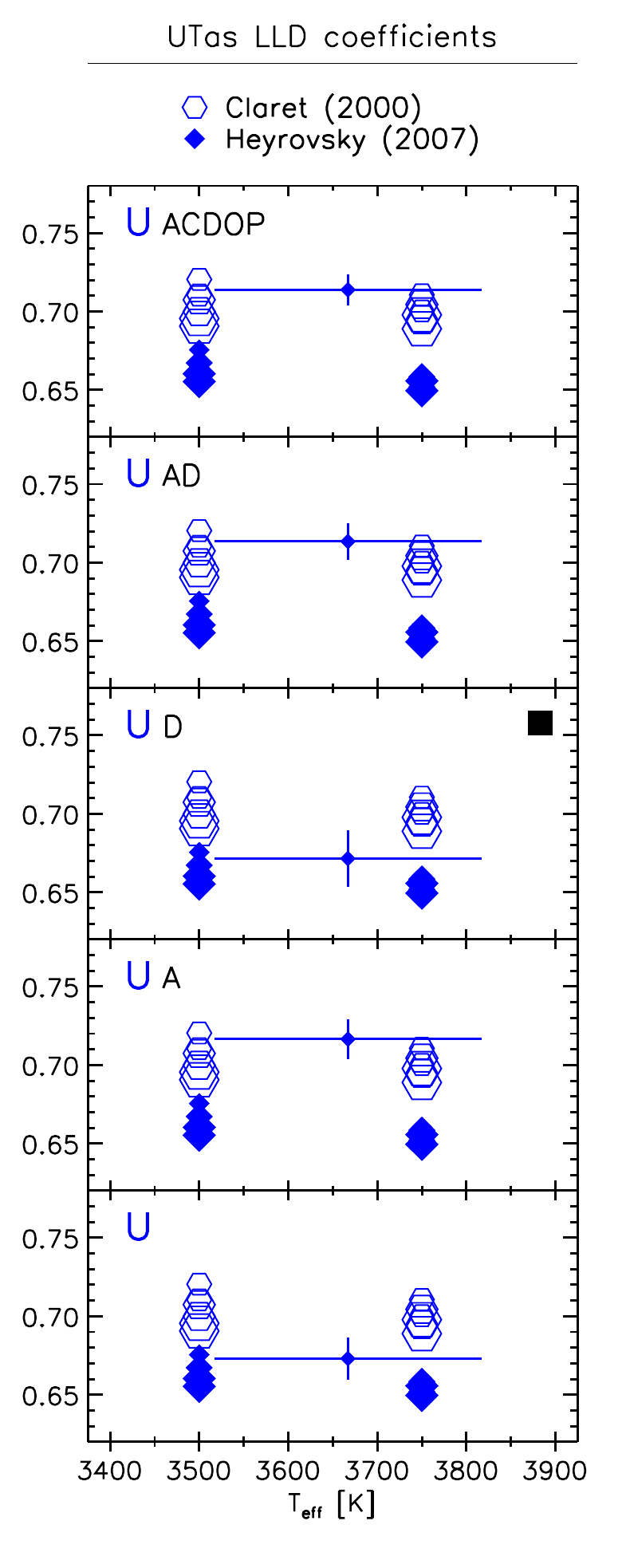}  
     \includegraphics[width=5.9cm, bb= 0 0 227 567]{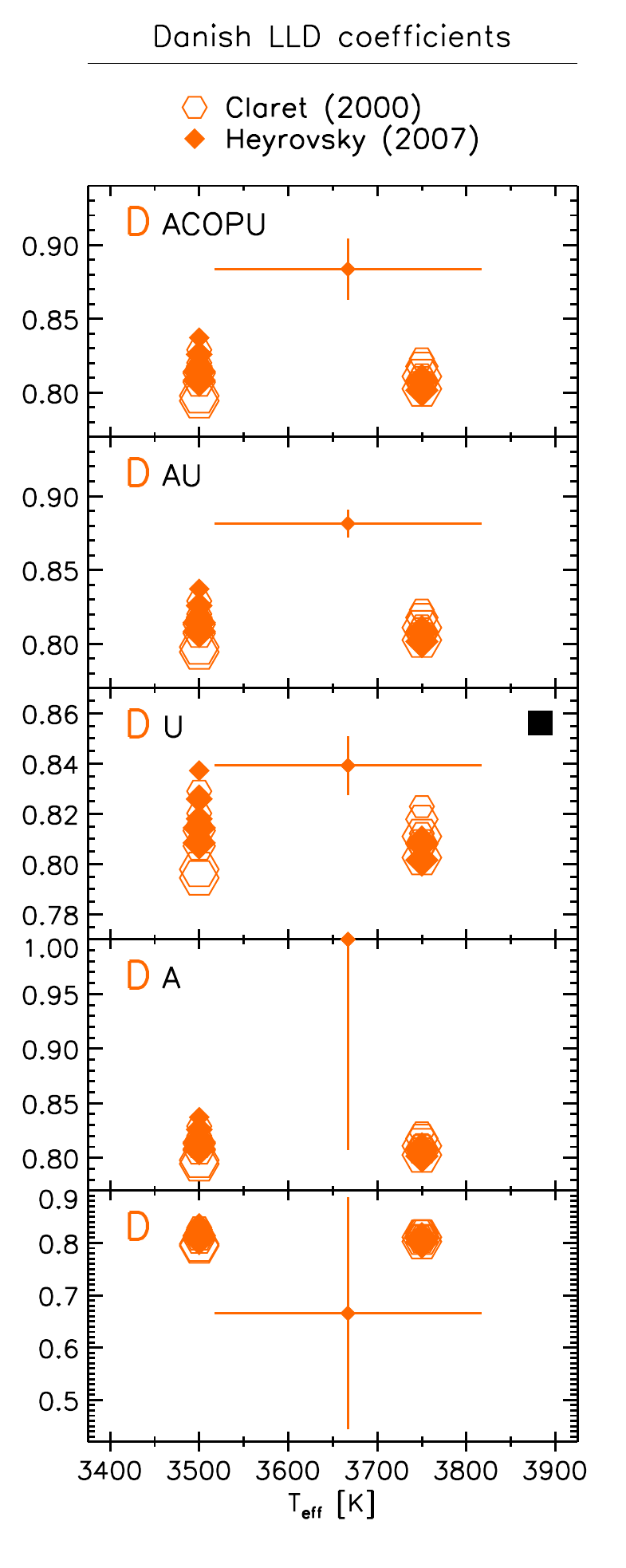} 
     \includegraphics[width=5.9cm, bb= 0 0 227 567]{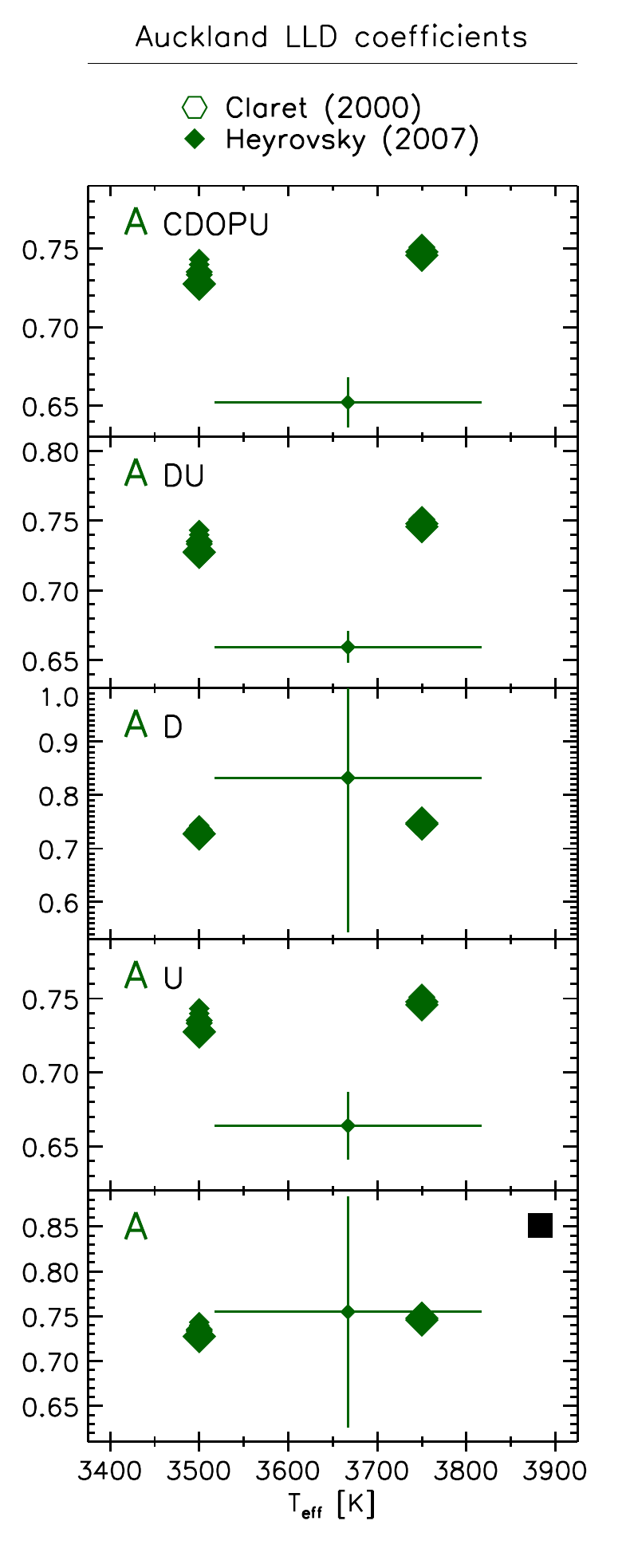}    
     \caption{
       Graphical representation of the linear limb-darkening measurements (crosses) for the three data
       sets with sensitivity to limb-darkening: UTas in the $I$-band, Danish in the $R$-band, and
       Auckland in a clear filter. 
       The open hexagons and the filled diamonds are the predictions from
       \citet{Claret2000} and \citet{Heyrovsky2007} linear limb-darkening (LLD) coefficients. 
       The fitting of the light curve is performed for different combinations of telescopes
       (same letter conventions as for \Tab{tab:fits}), and the results are discussed in 
       \Sec{sec:LDmesure}. The adopted measurements are those marked with black squares in the 
       upper right of the panels. }
     \label{fig:LLD-ac}
   \end{center}
 \end{figure*}

 \section{Linear limb-darkening discussion} \label{sec:LDmesure}
 
 As discussed in \Sec{sec:fitting}, three data sets have some sensitivity to limb-darkening: UTas
 ($I$-band), Danish ($R$-band) and Auckland (clear filter). The first question we address now
 is how the individual linear limb-darkening coefficients
 (LLDC) are affected by including or removing some of our data sets. 
 Indeed, our first step was to model every data set independently, and step by step to include
 other telescopes. We first noticed that there was a change in the LLDC values that depends on
 the added data sets. We therefore performed a detailed analysis to understand what could cause such
 variations, and to identify combinations of data sets that lead to correct
 LLD measurements. The results we are commenting are presented in \Fig{fig:LLD-ac}: 
 the three columns correspond to
 UTas, Danish, and Auckland, respectively, and the individual panels
 display the LLDC measurements for various combinations of data sets; the corresponding model parameters are given in
 \Tab{tab:fits}. In the figure and table, the letters A, C, D, O, P, U refer to the
 telescopes Auckland, CTIO Yale, Danish, OGLE, Perth, and UTas respectively.

UTas (U) clearly provides the best data set for LLDC measurements, since the
 data sample the whole LLD-sensitive region at the peak of the light
 curve, as well as its wings and baseline. 
On the other hand, modelling Danish (D) alone provides a very unrealistic fit, with large 
 error bars and very irregular MCMC correlations.
 To explain this, we recall that as mentioned in \Sec{sec:data}, the peak of the light curve was
   observed under bad weather conditions in La Silla, in particular
 the two consecutive data points around $t=3235.5$. Moreover, the data coverage is not optimal, because there are only two epochs 
   that cover the LLD-sensitive part of the light curve. 
   As a result, this poor coverage combined with some
   uncertainty in the data lead to a model that apparently looks better
   in terms of chi-squared, but cannot be trusted.
The last telescope with data sensitive to limb darkening is Auckland (A). We can fit the
 corresponding data alone and obtain a reasonable fit, but we obtain large error bars because the 
 photometric accuracy of the data is several times lower than for UTas, and furthermore, 
 the data taken during the source crossing are all located close to the limb, in
 the region of less sensitivity to limb darkening.
 We note that the LLDC we obtain is higher than UTas's, which
 is expected, because Auckland's clear filter is known to peak between red and infrared and LLDCs
 usually decrease towards the infrared.  

Starting from these models, we include different combinations of other data sets. If we base our
 analysis on the LLDC measurement from our best data set, UTas, then we find two distinct
 behaviours: either the UTas's LLDC is not displaced from the individual fit 
 ($a \simeq 0.67$, e.g. U or U+D) or is slightly modified
 ($a \simeq 0.71$, e.g. U+A or all telescopes).  
   Interestingly, when combining U and D
   data, the fit is stabilised for D. This is because
   the common fitting parameters ($\uo, \to, \tE, \rhoS$) are better
   constrained. 
   However, combining A with U data modifies the LLDCs compared to A
   and U modelled alone. This is also true when combining U+D+A or all
   telescopes. 
   That Auckland modifies UTas's LLDC (our best-suited data
   set for LLDC measurements) when the two data sets
   are combined, lead one to be careful about the interpretation of Auckland's
   LLDC (besides the large error bar on the LLDC).
  
 From this, we conclude that a precise measuring of LLDC requires a
 very careful study: first, one has to identify the data sets that can potentially provide a limb-darkening measurement with
 enough sensitivity, based on 
 the light-curve sampling as discussed in \Sec{sec:fitting}. 
 Then, one has to check whether the inclusion of
 additional data sets affects the results. Indeed, as we have shown for this microlensing event,
 adding more data sets to the light curve modelling can lead to two opposite effects:
 either the new data stabilise the fit and help obtain
 LLDCs for more data sets, or they perturb the LLDC measurements. The latter may happen if
 unknown systematic effects are affecting the data.
 For \event, the most reliable LLDCs for 
   UTas and Danish are obtained when these data sets are combined in
   the fit (U+D). No definitive conclusion can be
   safely drawn for Auckland LLDC, although its best estimation is
   likely to be obtained using A data alone. 
 The relevant measurements we discuss below are marked in \Fig{fig:LLD-ac} with a
 black square in the upper right of the corresponding panels.
 When the fit is performed using the formula of
 \cite{Heyrovsky2003}, we obtain similar results for 
 the combinations U+D and A: $a_{\rm U} = 0.655^{+0.010}_{-0.016}$, $a_{\rm D} =
 0.825^{+0.023}_{-0.022}$ and $a_{\rm A} = 0.751^{+0.083}_{-0.096}$.

In order to compare our measurements to linear limb-darkening predictions from atmosphere models,
 we use two sets of LLDCs computed from Kurucz's ATLAS models, for
 which hydrostatic equilibrium and LTE were assumed \citep[e.g.][]{Kurucz1992,Kurucz1994}. 
 The first set of LLDCs is taken from \citet{Claret2000},
 using the VizieR database, for the whole available range of 
 temperatures and $\logg$ compatible with \event's
 source star fundamental parameters (\Sec{sec:srceparam}); we assume a solar metallicity 
 to be consistent with our spectral analysis.
 The corresponding LLDCs are plotted 
 in \Fig{fig:LLD-ac} as thin, open hexagons. 
 We find twelve models that correspond to our requirements: two different temperatures ($3500$ and $3750$~K), three $\logg$
 ($1.0$, $1.5$ and $2.0$, plotted from smaller to bigger symbols) and for each configuration two
 microturbulent velocities ($1$ and $2$~km/s). The second set of LLDCs is plotted as filled diamonds, and
 correspond to coefficients computed using the interpolation method advocated by
 \citet{Heyrovsky2007}. These are computed for the same stellar parameters as before.
 
 Evidently our LLDC measurements agree very well with the predictions from atmosphere
 models. For UTas $I$, our measurement is compatible with both the predictions
 from \citet{Claret2000} and \citet{Heyrovsky2007}. For the Danish $R$ filter, the agreement is also
 very good, although our measurement is slightly larger than the
 prediction. For the Auckland
 clear filter, only \citet{Heyrovsky2007} predictions are available; but within the large error bars
 commented on previously, the data are fairly compatible with the predictions. 
 
 \section{PCA-based limb-darkening coefficients} \label{sec:PCA}

 Although stellar limb darkening is usually modelled by
 analytical laws, another option is to construct new bases of
 functions directly from model-atmosphere limb-darkening profiles. 
 In this section, we use a limb-darkening basis numerically
 constructed by principal component analysis (PCA and PCA LD in the
 following) for a set of given model atmosphere limb-darkening
 profiles, following \citet{Heyrovsky2003}. 

 \begin{figure}[!ht]
   \begin{center}
     \includegraphics[width=8cm]{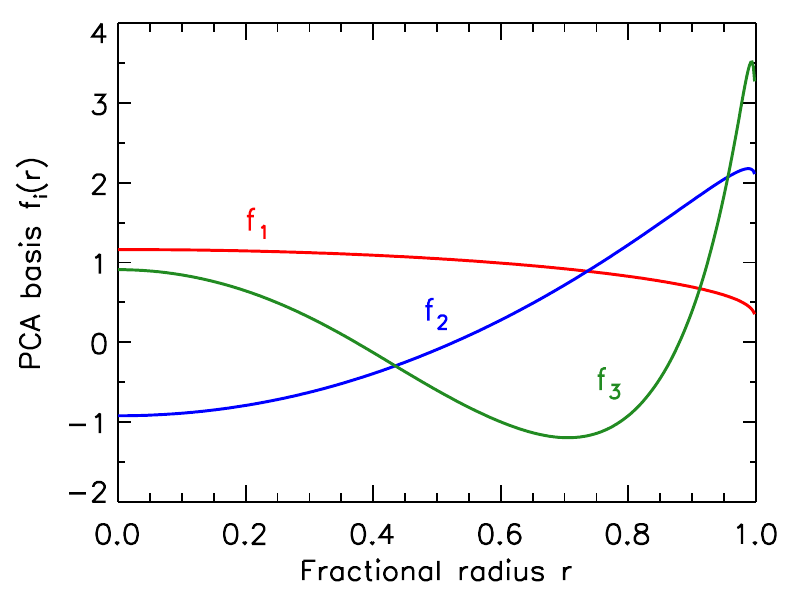}     
     \caption{     
     First three basis functions of the PCA decomposition, computed
     for a set of atmosphere models that match the stellar parameters of 
     \event's source star.}
     \label{fig:PCAB}
   \end{center}
 \end{figure}

 In this approach, the stellar intensity profile is expressed as 
 \begin{equation}
   I(r) = \sum_i \alpha_i\,f_i(r)\, ,
 \end{equation}
 in place of \Eq{eq:ldatm},
 where the $f_i (r)$ are the PCA basis functions and $\alpha_i$ 
 the corresponding coefficients.
 For our analysis we used a very general PCA basis constructed from the $BVRI$ profiles of the full Kurucz (1992) ATLAS model-atmosphere grid (see Heyrovsk\'y 2008 for details). 
 The resulting three first basis 
 functions, computed for a set of atmosphere models that match the stellar parameters of 
 \event's source star, are displayed in \Fig{fig:PCAB}.

 In the simplest case of a 2-term PCA LD law (the analogue of the analytical linear
 limb-darkening law, LLD), the relevant parameter that
 defines the shape of the star's brightness profile 
 is $\kappa \equiv \alpha_2/\alpha_1$. 
 With our choice of PCA basis, all possible profiles are obtained by varying  
 $\kappa$ in the range $[-0.162,\, 0.090]$, from the most peaked to the 
 flattest limb-darkening profiles.

\begin{figure*}[!ht]
   \begin{center}
     \includegraphics[width=5.9cm]{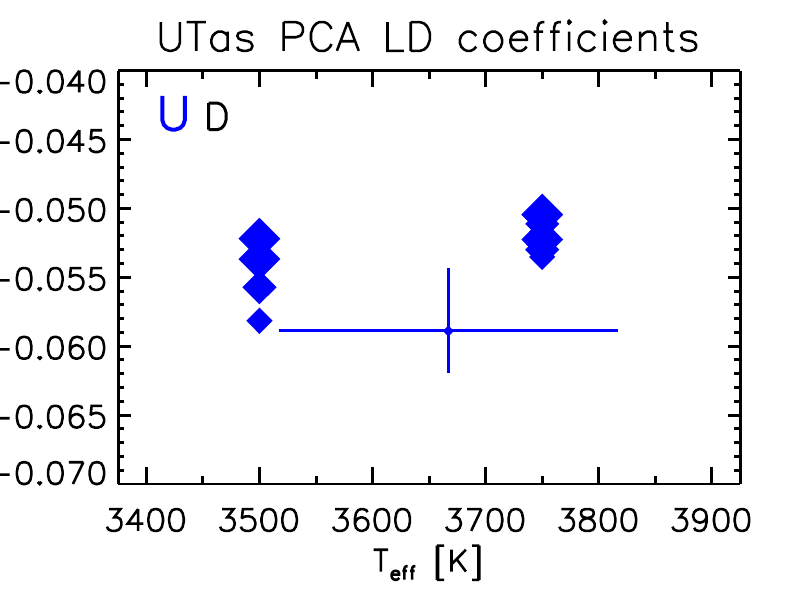}        
     \includegraphics[width=5.9cm]{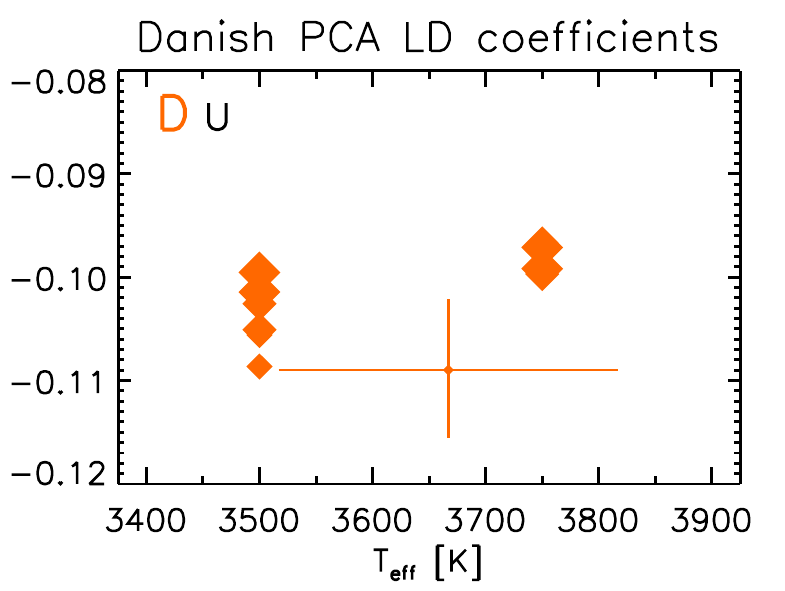}     
     \includegraphics[width=5.9cm]{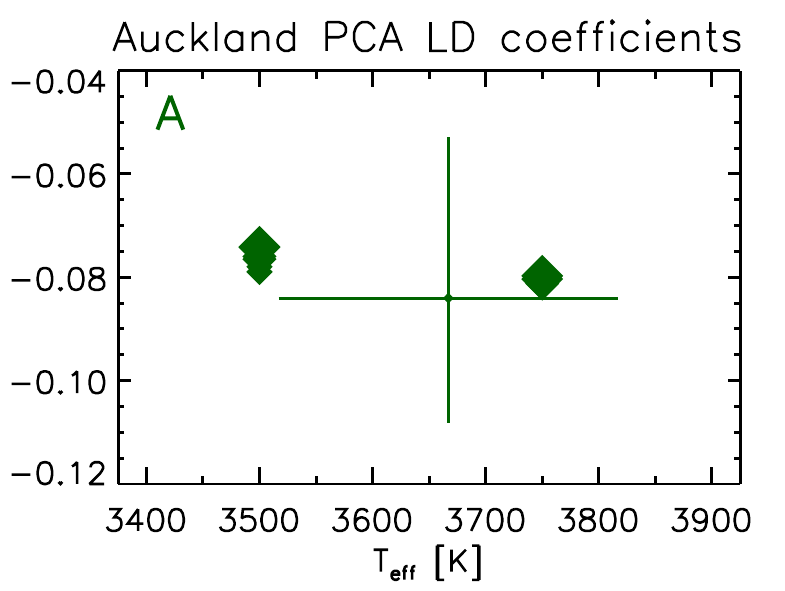}     
     \caption{
       PCA limb-darkening (PCA LD) coefficients $\kappa$ measured (crosses) and predicted
       \citep[diamonds,][]{Heyrovsky2008conf} using
       the 2-term PCA LD as explained in the text. Letters and colours have the same meaning as in
       \Fig{fig:LLD-ac}. }
     \label{fig:LLD-dh}
   \end{center}
 \end{figure*}

 We performed the OGLE 2004-BLG-482 analysis using the
 \cite{Heyrovsky2003} formalism for different combinations of data sets in a similar way as in \Sec{sec:LDmesure}. The results are presented in \Fig{fig:LLD-dh}
 for the combinations of data sets that were selected in the previous
 section (\Fig{fig:LLD-ac} panels with a black square in the upper right).

 As for the classical LLD law discussed in
 detail in the previous section, we find a very good
 agreement between model predictions and our measurements. This shows that PCA LD
 provides an interesting alternative to
 analytical models of stellar brightness profiles.
   For applications where LD is not fitted \citep[e.g.,][]{Jovi},
   it can be interesting to use PCA rather than LLD laws.
   On the other hand, the PCA LD law always depends on the
   set of selected model atmospheres. This could lead to discrepancies
   for instance if the parameter grid is too narrow. In addition, any
   PCA LD law reflects the physics included in the construction of the
   particular atmosphere model (e.g., variants of ATLAS, MARCS, or
   PHOENIX models), which may not be ideally suited for the studied
   star. In either of these cases, however, if the observational data
   are good enough, one may use the situation to one's own benefit. By
   trying different PCA LD laws and checking the quality of the fits
   and patterns in the residuals one can discriminate between
   different ``candidate'' model atmospheres. To summarize, in
   limb-darkening modelling LLD has the advantage of simplicity and
   analyticity, while PCA LD has the advantage of providing better
   accuracy and flexibility.
 
 \section{Conclusion} \label{sec:discussion}
 
 We have analysed \event, a relatively high-magnification single-lens microlensing event
 with notable extended-source effects, which was densely covered by our telescope networks.
 We derived precise limb-darkening measurements of the source star, a
 cool M giant, in particular in the $I$ and $R$ bands by combining the UTas and Danish data sets.
 No definitive conclusion could be made for Auckland data, 
 affected by unknown systematics that prevented the data to be well-fitted
 along with other data; however, when the Auckland data are fitted alone, the derived 
 limb-darkening agrees to model predictions, but with a large uncertainty.
  
 It provided us with the rather rare opportunity to directly test
 model-atmosphere limb-darkening predictions for the source star. This comparison was made possible 
 because we could obtain high-resolution UVES spectra at VLT at a critical time thanks to the short
 activation of a ToO programme at VLT, from which we could precisely estimate the star's fundamental
 parameters. The source typing has been confirmed with good precision by our photometric
 diagnostic based on a calibrated colour-magnitude diagram of the field.
 We have performed a very detailed modelling to evaluate the impact of including data sets in
 the modelling process, and provide new diagnostics for future work. 

 Very interestingly, the measured limb darkening agrees very well with model-atmosphere predictions, 
 both when considering linear limb-darkening profiles,
 or an alternative limb-darkening description
 based on a principal component analysis of ATLAS stellar atmosphere models. 
 From this study, where the precision has been pushed to a high level, we conclude that
 this late M giant does not suffer from any clear discrepancy between limb-darkening model predictions and
 measurements, which has been pointed out for earlier K giants. Although it is based on the
 observation of a single event, it is very likely that the conclusion
 can be extended to similar late M giants.

 \begin{acknowledgements}
   We express our gratitude to the ESO staff at Paranal for reacting 
   at short notice to our UVES ToO request. 
   We thank David Warren for 
   financial support for the Mt~Canopus Observatory. The OGLE project 
   is partially supported by the Polish MNiSW grant N20303032/4275. 
   MZ acknowledges a partial support of the Polish
   Research Grant N N203 2738 33. 
   AC acknowledges travel support on the French CNRS/ANR grant
   HOLMES. DH was supported by Czech Science Foundation grant GACR 
   205/07/0824 and by the Czech Ministry of Education project 
   MSM0021620860. This publication makes use of data products from the 
   2MASS and DENIS projects, as well as the SIMBAD database, Aladin and 
   VizieR catalogue operation tools (CDS Strasbourg, France).
   CH acknowledges the support by Creative Research Initiative Program
   (2009-008561) of Korea Science and Engineering Foundation. BGP
   acknowledges the support by Korea Astronomy and Space Science
   Institute. We thank Subo Dong for his comments on the analysis.  
 \end{acknowledgements}
 
 \bibliographystyle{aa}
 \bibliography{12007OB482}
 \end{document}